\documentclass[twocolumn,nofootinbib,amsmath,prd,aps,superscriptaddress,tightenlines,preprintnumbers]{revtex4}

\pdfoutput=1

\usepackage{amsmath}
\usepackage{amssymb}
\usepackage{graphicx}
\usepackage{color}
\usepackage{tikz}
\RequirePackage[normalem]{ulem} 
\RequirePackage{color}\definecolor{RED}{rgb}{1,0,0}\definecolor{BLUE}{rgb}{0,0,1} 

\begin{document}

\newcount\hour \newcount\minute
\hour=\time \divide \hour by 60
\minute=\time
\count99=\hour \multiply \count99 by -60 \advance \minute by \count99
\newcommand{\mydate}{\ \today \ - \number\hour :00}
\newcommand{\andre}[1]{\textbf{\color{red} #1}}
\newcommand{\juan}[1]{\textbf{\color{blue} #1}}
\newcommand{\riley}[1]{\textbf{\color{purple} #1}}
\newcommand{\lb}{\texttt{laidbax}}
\newcommand{\pax}{\texttt{PaX}}
\renewcommand{\wr}{\texttt{wimprates}}
\title{A semi-supervised approach to dark matter
	searches in direct detection data with machine learning}

\def\coepp{ARC Centre of Excellence for Dark Matter Particle Physics, Department of
	Physics, University of Adelaide, Adelaide, South Australia 5005, Australia}

\author{Juan Herrero-Garcia}
\email{juan.herrero@ific.uv.es}
\affiliation{Departamento de F\'isica Te\'orica and IFIC, Universidad de Valencia-CSIC, \\ C/ Catedr\'atico Jos\'e Beltr\'an, 2, E-46980 Paterna, Spain}

\author{Riley Patrick}
\email{riley.patrick@adelaide.edu.au}

\affiliation{ARC Centre of Excellence for Dark Matter Particle Physics, Department of
	Physics, University of Adelaide, Adelaide, South Australia 5005, Australia}

\eprint{IFIC/21-39}
\author{Andre Scaffidi}
\email{andre-joshua.scaffidi@to.infn.it}

\affiliation{Istituto Nazionale di Fisica Nucleare, Sezione di Torino, via P. Giuria 1, I–10125 Torino, Italy}
\begin{abstract}
The dark matter sector remains completely unknown. It is therefore crucial to keep an open mind regarding its nature and possible interactions. Focusing on the case of Weakly Interacting Massive Particles, in this work we make this general philosophy more concrete by applying modern machine learning techniques to dark matter direct detection. We do this by encoding and decoding the graphical representation of background events in the  XENONnT experiment with a convolutional variational autoencoder. We describe a  methodology that utilizes the `anomaly score' derived from the reconstruction loss of the convolutional variational autoencoder as well as a pre-trained standard convolutional neural network, in a semi-supervised fashion. Indeed, we observe that optimum results are obtained only when both unsupervised and supervised anomaly scores are considered together.  A data set that has a higher proportion of anomaly score is deemed anomalous and deserves further investigation.  Contrary to classical analyses, in principle all information about the events is used, preventing unnecessary information loss. Lastly, we demonstrate the reach of learning-focused anomaly detection in this context by comparing results with classical inference, observing that, if tuned properly, these techniques have the potential to outperform likelihood-based methods.

\end{abstract}

\maketitle

\section{Introduction} \label{sec:intro}

The origin of dark matter (DM) of the Universe is an unresolved mystery, despite the unprecedented experimental effort of the last decades, in particular in the context of Weakly Interacting Massive Particles (WIMPs). One of the most relevant type of searches is looking for nuclear recoils in underground detectors, known as direct detection (DD)~\cite{Goodman:1984dc}. Already now, and in the coming years, several xenon (PandaX~\cite{Cui:2017nnn}, XENONnT~\cite{Aprile:2014zvw}, LZ~\cite{Akerib:2015cja} and ultimately DARWIN~\cite{Aalbers:2016jon}) and argon (DEAP-3600~\cite{Fatemighomi:2016ree}, 
DarkSide and Argo~\cite{Aalseth:2017fik}, ArDM~\cite{Calvo:2016hve}) ton-scale experiments will improve significantly the sensitivity, see also Refs.~\cite{MarrodanUndagoitia:2015veg,Schumann:2019eaa} for the experimental status of DD experiments. In addition, the DAMA annual modulation signal (see Ref. \cite{Freese:2012xd} for a review) is being tested in a model-independent with several experiments using sodium iodine (COSINE~\cite{Adhikari:2019off}, ANAIS~\cite{Amare:2019jul}, COSINUS~\cite{Angloher:2016ooq} and SABRE North/South~\cite{Shields:2015wka,Froborg:2016ova}). 

Machine learning (ML)  approaches are a growing area of research in the physics community. Indeed, their application in DM phenomenology is starting to become more pronounced. 
In Ref.~\cite{Zhang:2019ryt} a two-phase CNN architecture undertaking first classification and then regression was used to map 3D galaxy distributions to its underlying DM distribution. In Ref.~\cite{Lucie-Smith:2019hdl} one can see gradient boosted trees used to model DM halo formation. Similarly, in Ref.~\cite{Bernardini:2019bmd} a deep CNN architecture is used to model halo formation.  In the collider physics context many studies have been published in recent years that attempt to parametrize the concept of anomalousness in data using unsupervised machine learning - see Refs.~\cite{Farina:2018fyg,Heimel:2018mkt,Hajer:2018kqm,Kuusela:2011aa,cerri2019variational,knapp2020adversarially,Andreassen:2020nkr,Nachman:2020lpy,Collins:2019jip,Dery:2018dqr,Collins:2018epr,Otten:2019hhl, Khosa:2020qrz, vanBeekveld:2020txa}.  These studies employ a variety of different methods in a number of different ways; of particular relevance to this paper are Refs.~\cite{cerri2019variational} and \cite{vanBeekveld:2020txa} which employ a variational autoencoder (VAE) on simulated events of the CMS and ATLAS detectors respectively, as well as Ref.~\cite{Khosa:2020qrz} where an algorithm dubbed ``anomaly awareness'' is proposed that utilizes a single supervised CNN architecture to identify anomalous events in boosted hadronic jet structure. In this work we show how  using more than one of these such architectures can yield optimal results.

 In Ref.~\cite{Khosa:2019qgp}  a convolution neural network was applied to XENON1T TPC detector response images of WIMP signal events (nuclear recoils) and background events (electron recoils) achieving a classification accuracy of above $\sim95\%$. In this study, we exploit a pre-trained convolutional neural network with a similar architecture to boost the overall anomaly awareness of our unsupervised model. In anomaly-based unsupervised analyses, the key is to train a model to successfully capture the information contained within a background data set such that data points that don't conform to the properties of the background can be identified as `anomalous'. In this work we apply this concept to DM direct detection. For this goal we train a convolutional variational autoencoder (CVAE) on a data set of composite images containing the detector response and S1/S2 signals of simulated electron recoil events from the XENONnT DM detector. We then apply the trained CVAE to simulated data sets with a WIMP signal present, so that the reconstruction loss for these events is on average higher, serving as a measure of anomalousness. We then utilize a previously trained convolutional neural network (a slightly tuned version of that from Ref.~\cite{Khosa_2020}), capable of classifying background and signal at $\sim 98\%$ accuracy to construct an even more optimized `anomaly score', observing that its asymptotic properties give the best anomaly identifying power.

     While purely unsupervised methods will likely never outperform supervised classification methods they are still powerfully motivated by their ability to identify events which are not background-like in a way that is agnostic to all signal models. In the DM context where the parameter space of possible models is extremely large, the ability to find any anomalous events is helpful in motivating future studies, as well as identifying unknown systematic errors.

This paper is structured as follows. In section ~\ref{sec::dd} we briefly review the ingredients involved in the DM direct detection event rate. In Section~\ref{sec:TPC} we discuss the XENON experimental TPC. We outline the generation of the data for the analysis in Section~\ref{sec:data}. Section~\ref{sec:cvae} provides background on the concept of a variational autoencoder and convolution.  In Section~\ref{sec:architecture} we present our CVAE architecture and discuss the training. In Section~\ref{sec:analysis_and_results} we present our analysis pipeline, including an overview of the anomaly score functions utilized for anomaly detection.  Here we show the results demonstrating our methods capability of significantly detecting anomalies and forecasting sensitivity, as well as a caparison of projected sensitivity compared to a traditional likelihood based technique.   We then conclude in Section~\ref{sec:conc}.

\section{Dark matter direct detection}
\label{sec::dd}
In this section we briefly review the direct detection of DM by measuring the nuclear recoil due to its scattering in underground detectors, see for instance Ref.~\cite{Cerdeno:2010jj}. We focus on elastic scattering of DM with mass $m_\chi$, with a detector consisting of one type of nucleus of mass $m_A$, mass number $A$ and atomic number $Z$. The differential event rate per unit detector mass is given by
\begin{equation}
\frac{\mathrm{d}\mathcal{R}}{\mathrm{d}E_R} = \frac{\rho_\chi}{m_\chi m_A} \int_{v_{\rm min}}  \frac{\mathrm{d}\sigma_{\chi A}}{\mathrm{d}E_R}\, v \,f_{\rm det}(\hat{\bf{v}})\, \mathrm{d}^3v\,,
\label{eq:difrate}
\end{equation}
where $\rho_\chi$ is the local WIMP density,  $f_{\rm det}(\bf{v})$ is the WIMP velocity distribution in the detector rest-frame and $v_{\rm min}$ is the minimum WIMP velocity required to produce a recoil, above the experiment's threshold, of energy $E_R$,
\begin{equation}
v_{\rm min} = \sqrt{\frac{m_A\,E_R}{2\mu^2_{\chi A}}}\,,
\end{equation}
with $\mu_{\chi A} = m_\chi m_A/(m_\chi + m_A)$ being the DM-nucleus reduced mass. For contact interactions, the differential scattering cross section reads
\begin{equation} \label{eq:diffcs}
\frac{\mathrm{d}\sigma_{\chi A}}{\mathrm{d}E_R} = \frac{m_A}{2\mu_{\chi A}^2 v^2}\big[\sigma_{\rm SI}F^2_{\rm SI}(E_R) + 
\sigma_{\rm SD}F^2_{\rm SD}(E_R)\big]\,,
\end{equation}
where $F_{\rm SI\,(SD)}$  are the
spin-independent (SI) and spin-dependent (SD) form factors, and $\sigma_{\rm SI,SD}$ are the corresponding WIMP-nucleus cross 
sections at zero momentum transfer~\cite{Engel:1992bf},
\begin{align} \label{eq:csSISD}
\sigma_{\rm SI}&= \frac{4\mu_{\chi A}^2}{\pi}\left(Z\,f_p +  (A-Z)\,f_n   \right)^2\,, \\
 \sigma_{\rm SD}&= \frac{32}{\pi}G^2_F\mu_{\chi A_2}\frac{J+1}{J}\left[a_p\langle S_p\rangle + a_n\langle S_n \rangle  \right]^2 \,.
\end{align}
Here $G_F$ is Fermi's constant, $f_{p\,(n)}$ is the WIMP-proton (neutron) SI coupling, $a_{p\,(n)}$ is the 
WIMP-proton (neutron) SD  coupling, $J$ the angular momentum of the target nucleus, and $\langle S_{p\,(n)}\rangle$  the spin content of the proton (neutron).

Assuming only SI interactions, the differential event rate simplifies to
\begin{equation}
\frac{\mathrm{d} \mathcal{R}}{\mathrm{d}E_R} =  \frac{\rho_\chi\,\sigma_{\rm SI}}{2m_\chi \mu_{\chi A}^2} 
F^2_{\rm SI}(E_R) \int_{v_{\rm min}} \frac{f_{\rm det}(\hat{\bf{v}})}{v}\mathrm{d}^3v \,.
\label{eq:dRdEr}
\end{equation} 
In this paper, we consider SI interactions with xenon nuclei (in the following we drop the ${\rm SI}$ label). We consider a Standard Halo Model. We will make use of  \wr~\cite{wr} to generate the DM event rate (see this reference for the values used in this analysis).

\section{TPC Event Reconstruction in the XENON Experiment} \label{sec:TPC}

In this work we simulate background and signal events for the XENONnT experiment located in the Gran Sasso National Laboratory. XENON1T has been running and it is being superseded by XENONnT. The latter uses a dual-phase time projection chamber with 8 tons of natural xenon, of which around 6 tons constitute its fiducial volume. It consists of a liquid xenon target with a gaseous phase on top, with an applied electric field throughout the detector. The detector has two arrays of around 250 photomultipliers (PMTs) in each of the top and bottom layers. It is embedded in a water tank, with muon and neutron vetos \cite{XENON:2020kmp,Antochi:2021wik,Zhang:2021atm}. 

The main electronic background comes from gamma rays, being generated in the cryostat and in the PMTs, as well as the radioactive decay of  $^{222}$Rn and $^{85}$Kr from the detector material. The detection technique makes use of both the prompt scintillation signal or S1, and the scintillation created from the drifted electrons due to the electric field (ionisation signal) or S2. Fiducialisation of the detector is possible thanks to a full 3D position reconstruction by using both the time delay between S1 and S2 signals, as well as the number of photons seen by each PMT (e.g. the hitpatterns). 

\subsection{Model independence and supervised classification }
\label{sec::model_indep}
The key discrimination parameter to distinguish background (electronic recoils in this context\footnote{Recently the XENON1T Collaboration reported an excess in electron recoils over known backgrounds at low energies~\cite{Aprile:2020tmw,Athron:2020maw}. In this work we consider the standard case of a signal produced by WIMPs scattering on nuclei.}) from signal (nuclear recoils) is the S2/S1 ratio, which is larger for electron recoils (ER) than for nuclear recoils (NR). For our purposes, it is important to emphasize that this constitutes the main discriminating power between NR and ER, and therefore the classification accuracy achievable by the neural network should be independent of the DM properties (i.e. the DM mass). We note, however, that on an event-by-event basis, the fundamental detector response to a NR vs ER is different, and therefore supervised classification techniques can be used to discriminate between signal and background, as was observed in Ref.~\cite{Khosa:2019qgp}. A key result we found whilst validating the results of this study was that such supervised methods work exceptionally well for background rejection \textit{regardless} of the DM mass and cross-section as well, since the fundamental property being learned by the model is the NR and ER detector response. In this sense, the supervised classifier discriminates between ER and NR completely model-independently\footnote{To clarify, here `model-independent' refers to the DM model (eg. mass, couplings, etc.). On an event-by-event basis, the supervised classifier discriminates solely among ER and NR, that is, anything that does not classify as ER should then be considered anomalous (non-background) physics.}.  In this study, we focus on the generalized implementation of semi-supervised machine learning for this task, focusing on training the model on ER background-only. For any given data-set, any non-ER event will be labelled as `anomalous' when passed through the network.    It should be noted that in a real experiment there are other sources of NR background (radiogenic neutrons, wall leakage etc.) \cite{Khosa:2019qgp}. We do not consider these in this study since in low background detectors these backgrounds are thoroughly suppressed relative to WIMP signal in the XENONnT experiment \cite{Khosa:2019qgp}. 

\section{Data Generation} \label{sec:data}
To generate the image event samples used for this analysis, we primarily follow the methodology of Ref.~\cite{Khosa:2019qgp} and point readers to their \texttt{github} repository~\cite{lucygithub} for detailed information about event generation. The ER background spectra are contained within \wr\, and are empirically determined using the polynomial fit detailed in Ref.~\cite{Aalbers:2018mfc}.\footnote{It should be noted that the ER model used in \lb\,  is not the official model	XENON collaboration, which  includes several more systematic uncertainties.}   These spectra are then parsed to \lb~\cite{lb}, which performs the MC generation of events, effectively sampling from the given energy spectrum and producing detector observables for each event.  To generate the NR samples that would correspond to WIMP signal,  we utilize  \wr~\cite{wr} to generate energy spectra for an ensemble of WIMP masses in the range 10-1000 GeV. We use a spin-independent interaction cross-section of $\sigma = 10^{-45} \mathrm{cm}^{-2}$. As discussed in Sec.~\ref{sec::model_indep}, given that the classification analysis is done on an event-by-event basis, the only relevant property of these images is that they correspond to NR. The coupling strength of the WIMP-nucleon interaction as well as the mass of the particle is irrelevant and only serves to determine the overall expected number of events given some exposure. The cross-section therefore simply determines the weights for the Monte Carlo (MC) event sampling. This weighting will effect the projected sensitivity as discussed in Sec.~\ref{sec:forecast}.

 The physically relevant  variables taken for the analysis are recoil type (NR or ER), and corresponding S1, S2 signals, as well several auxiliary parameters. The model output from \lb\, is then parsed into \pax\,  (Processor for Analyzing XENON) ~\cite{pax}, which generates images of the  largest S1 and S2 peaks as well as the hitpatterns for the top and bottom PMTs. Ref.~\cite{Khosa:2019qgp} concluded that classification accuracy was maximized when both the S1 and S2 hitpatterns and the largest S1 and S2 peak graphs where displayed in the images.  Generating images in this manner meant that only information unique to each WIMP or ER event were present, effectively increasing the robustness of classification. We assume that, irrespective of the architecture used, the composite hit-pattern + peak images yield the best classification accuracy. Examples of an ER background event  (top) and a WIMP event (bottom) are shown in Fig.~\ref{fig:ER-example}.

\subsection{Image pre-processing}
\begin{figure}
\centering
\includegraphics[scale=0.2]{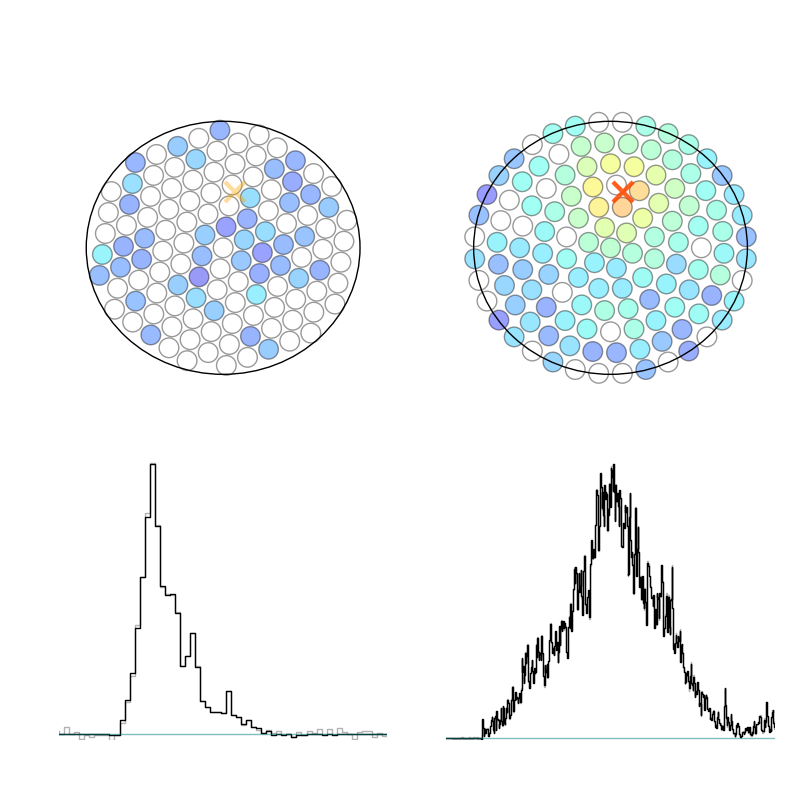}
\includegraphics[scale=2.13]{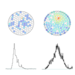} \\
\includegraphics[scale=0.2]{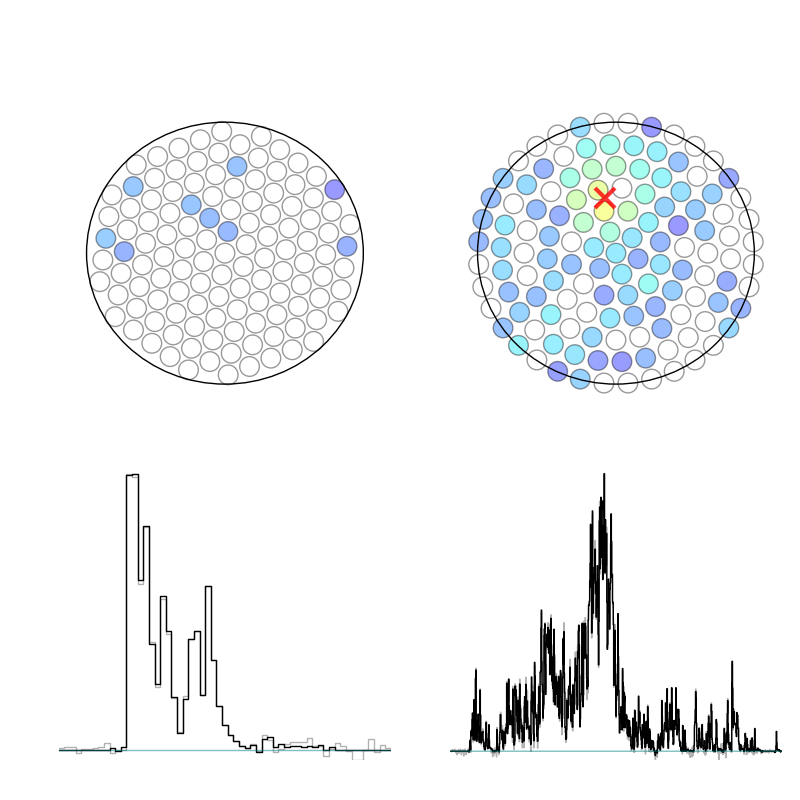}
\includegraphics[scale=2.13]{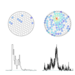}
\caption{\label{fig:ER-example} \textbf{Top:} An example of an 800$\times 800$ ER event image before (left) and after (right) reducing the resolution to $75\times 75$. \textbf{Bottom:} Same but for a NR (signal) event. The colour scale on the hit-pattern images denotes photon counts in the PMT arrays at the bottom (left) and top (right) of the detector. The largest S1 (left) and S2 (right) signal peaks are shown on the bottom half of the images. }
\end{figure}

For the analysis we generate an ensemble of 20000 NR and 20000 ER images. To make training manageable for the architecture presented in this work, the images are scaled down from their original $800\times 800$ resolution to $75\times75$ pixels. After validating the results from Ref.~\cite{Khosa:2019qgp} with the supervised convolutional neural network, we find  that this resolution gives optimal classification accuracy as opposed to $64\times64$ which is often adopted. We therefore also take an input image resolution of $75\times75$ for our CVAE architecture. In Fig.~\ref{fig:ER-example} we show examples of an ER and NR event both before (left) and after (right) lowering the image resolution.

\section{Convolutional Variational Autoencoders}
\label{sec:cvae}
Autoencoders are a type of neural network that attempt to return output that is as similar to their input as possible. This similarity is parameterized by a `reconstruction loss'. The smaller the loss, the more similar the output is to the input. The aim is to produce an encoding of the input data for the purposes of dimensional reduction, feature learning and produce generative models \cite{bank2021autoencoders,Schmidhuber_2015}.

Autoencoders consist of three main parts: an encoder, a latent space, and a decoder. The data inputs  enter into the encoder where they are condensed down into the latent space. This can be thought of as the encoding of the input information to a lower dimensional representation. The decoder then expands this lower dimensional representation of the data back out to the same dimensions as the inputs. For a sample of $N$ inputs, the difference between the input image and the output image is defined as a reconstruction loss, commonly a sum of squared errors,
\begin{align}
	L = \frac{1}{N}\sum_i^N(x_i-y_i)^2\,,
	\label{eqn:me}
\end{align}
where $x$ is the input vector and $y$ is the output vector.

Variational Autoencoders (VAEs) differ from basic autoencoders in that before the latent space vector in an autoencoder we insert two vectors, a vector of means  $\boldsymbol{\mu}$ and a vector of variances $\boldsymbol{\sigma^2}$  of normal distributions $\mathcal{N}(\boldsymbol{\mu},\boldsymbol{\sigma^2})$. Rather than attempting to encode the input vector onto a lower dimensional vector of real numbers, we are now attempting to encode the input information onto a set of Gaussian distributions which when sampled from provide an output vector that is as similar to the input vector as possible. These distributions serve to ``cover'' the latent space with smooth distributions which is a more faithful representation of smooth and continuous data \cite{Kingma_2019}.

Theoretically when a VAE with a standard reconstruction loss from  Eq.~\eqref{eqn:me}  is trained, it could simply learn Gaussians with extremely small variances (delta functions) or extremely disparate means (poor coverage of the latent space) and thus the VAE would effectively be a standard autoencoder and all benefits would be lost. Thus we regularize the reconstruction loss by penalizing distributions that are different from standard normal distributions. This is done by addition of the Kullback-Leibler divergence to the reconstruction loss,
\begin{align}
	L = \frac{1}{N}\sum_{i=0}^N(x_i-y_i)^2+\beta L_\text{KL}\,,
	\label{eqn:recon_loss}
\end{align}
where, for a latent space of $K$ standard normal distributions we have the special case of:
\begin{align}
	L_\text{KL} = \sum_{j=0}^K \left[\sigma^2_j+\mu^2_j-\log(\sigma_j)-1\right]\,,
\end{align}
which is clearly minimized for $\sigma_j = 1$ and $\mu_j = 0$. The $\beta$ factor is a parameter of $O(1)$ that scales the overall importance of the KL divergence to the reconstruction loss~\cite{Higgins2017betaVAELB}. The default $\beta = 1$ is the definition for a standard VAE, however we find optimal performance for $\beta = 0.5$ in this work.

A \textit{Convolutional} Variational Autoencoder (CVAE) is a VAE architecture which utilizes convolutional layers to optimize feature extraction from images and reduce the dimensionality of inputs. For a full discussion of image convolution in the context of machine learning see Ref.~\cite{dumoulin2018guide}.

\section{CVAE architecture and training}
\label{sec:architecture}
\subsection{Architecture}

\begin{figure*}
	\centering
	\includegraphics[width=\textwidth]{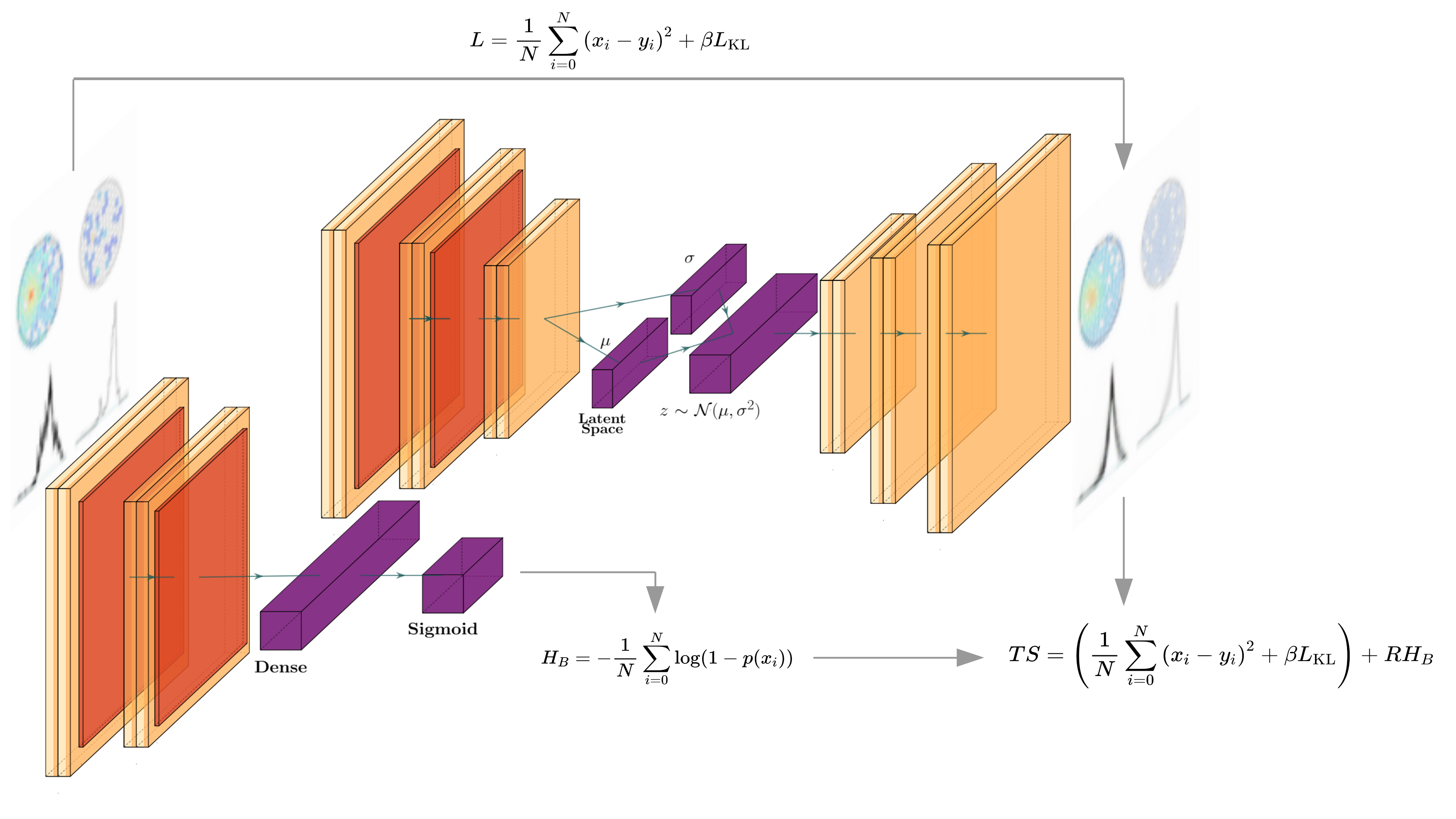}
	\caption{\label{fig:autoencoder-example} Our semi-supervised analysis pipeline including network architectures for the CVAE and CNN classifiers. Both networks are trained independently on the same training sets, and tested on the same testing sets. \textbf{Top:} CVAE architecture. The encoder is comprised of three convolutional layers with LeakyRelu activation (yellow), with the first two followed by pooling layers (red). These are then flattened onto a latent space consisting of 256 normal distributions each parameterized by their own means and variances. Samples $\mathbf{z}\sim \mathcal{N}(\boldsymbol{\mu},\boldsymbol{\sigma}^2)$ are then drawn from these distributions into a dense layer and then re-broadcast through the decoder using three transposed convolutional layers back into the original image dimensions on which the reconstruction loss (shown in Eq.~\ref{eqn:recon_loss}) is calculated between the input and output images (depicted via an arrow). The input and output images shown are actual realizations from our trained CVAE.  
		\textbf{Bottom:} The  simple CNN architecture that we used was primarily motivated by the results of Ref.~\cite{Khosa:2019qgp} and consists of two convolutional and pooling layers which are then flattened into a single dense output layer with a sigmoid activation function. This is used to calculate the binary cross-entropy in Eq.~\ref{eqn:cross_ent}, which will be larger for NR-like images and smaller for ER-like images. Both the reconstruction loss of the CVAE and the binary cross entropy from the CNN are then fed into the anomaly function (Eq.~\ref{eqn:TS}), which is then used to do inference.   
}      
\end{figure*}

\begin{figure}[ht]
	\centering
	\includegraphics[scale=0.62]{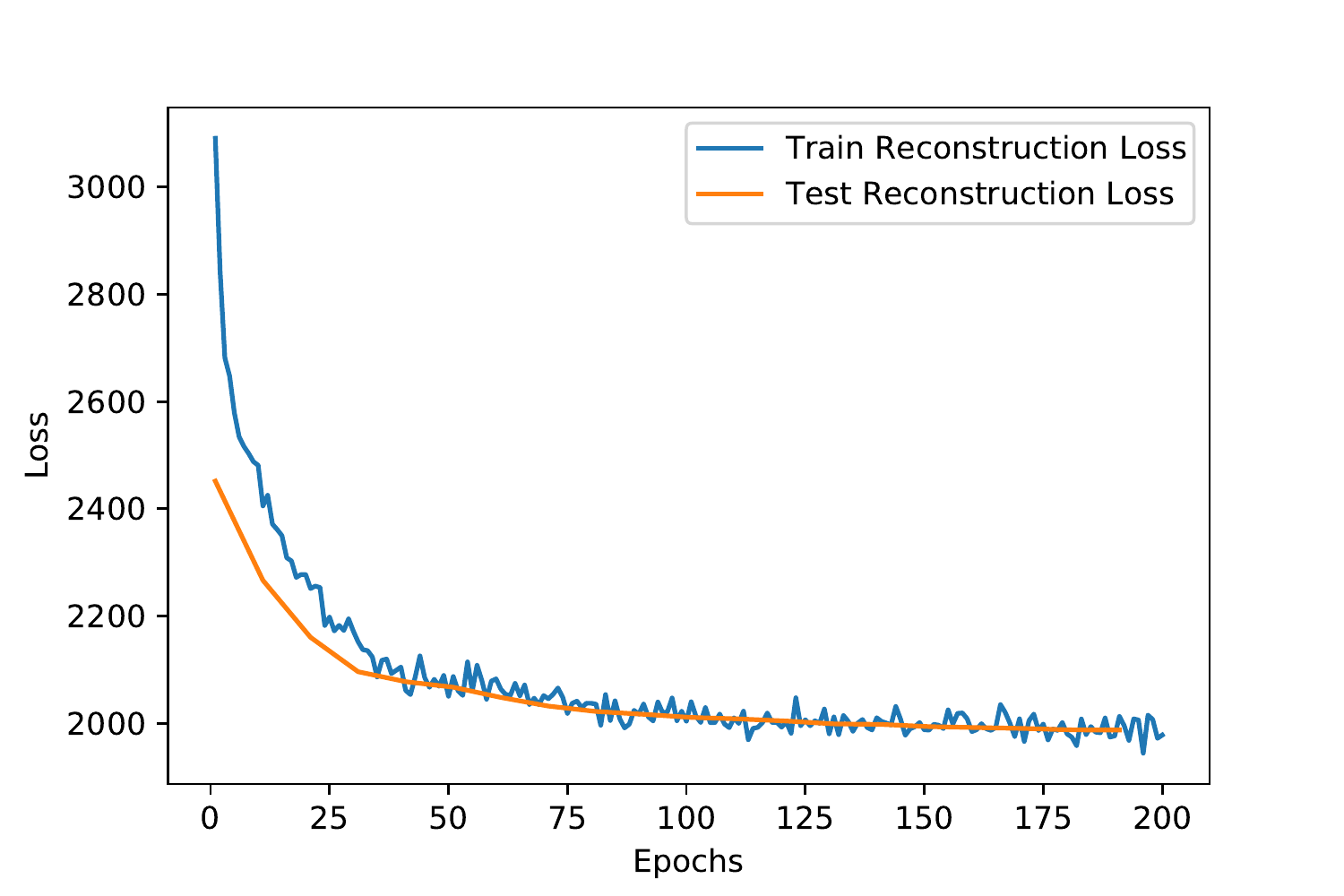}
	\caption{\label{fig:cvae-reco-loss-epoch} The reconstruction loss per epoch for the training and testing sets for the CVAE.}
\end{figure}
The CVAE architecture can be seen at the top of  Fig.~\ref{fig:autoencoder-example} and was implemented in \texttt{Tensorflow 2}~\cite{abadi2016tensorflow}. The network consists of a $75\times 75\times 3$ input layer, followed by three convolutional layers each with $128$ filters, $3\times3$ kernels and stride lengths of $3$. Each convolutional layer uses LeakyRelu activation with $\alpha= 0.05$, motivated by the hyper-parameter tuning undertaken in Ref~\cite{Khosa_2020}. The first two convolutional LeakyRelu layers are followed by pooling layers with dimensions $2\times2$ and $1\times 1$ respectively. The last convolutional layer is flattened and encoded onto latent space which consists of $512$ nodes, corresponding to $256$ means and $256$ variances, before being decoded via a reflection of the encoder using convolution transpose with ``same'' padding.

\subsection{Training and testing on background data}
\label{sec:training}
\begin{figure}[ht]
	\centering
	\includegraphics[scale=.55]{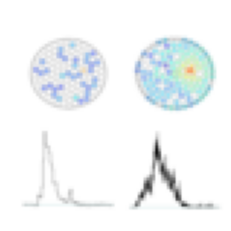}\includegraphics[scale=.55]{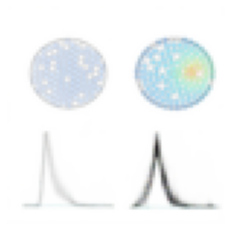}
	\caption{\textbf{Left:} Real ER image from the test set passed to the trained CVAE. \textbf{Right:} Output of the CVAE. \label{fig:sample_from_cvae}}
\end{figure}
We train the network for $200$ epochs on $16000$ ER event images in mini-batches of $100$ and test on $4000$. The training and testing reconstruction loss per epoch can be seen in Fig.~\ref{fig:cvae-reco-loss-epoch}. Both training and testing appears to converge with similar performance and overtraining  does not appear to be an issue.

Once trained, the CVAE should produce generative examples of ER images. This is demonstrated in Fig.~\ref{fig:sample_from_cvae}, where on the left we show a real ER image taken from the testing set and on the right we show the trained CVAE's reconstruction. One can note all primary features of the image are preserved, for example the general shape of the S1/S2 distributions in the bottom half, as well as the colour palate and general topology of the PMT hit patterns.  

Before continuing, we briefly draw attention to another well known use for variational autoencoders - that of latent space sampling. As the latent space is a lower dimensional representation of the properties of the input dataset by sampling from the latent space randomly, one may generate `new' events. This is an on-going and promising area of research, see for example Ref.~\cite{Otten:2019hhl}, as it is far more computationally efficient to sample from a latent space of a VAE than to perform costly Monte Carlo. In the case of this study the input data are images, and the pixel information of these images presents an extremely large input space which makes latent space sampling quite difficult. However, using generative models in place of Monte Carlo when the model utilizes raw detector data as inputs is an interesting avenue which we leave for future research. 

\section{Analysis and Results}
\label{sec:analysis_and_results}

\subsection{Anomaly detection}
\label{sec:forecast}

The analysis pipeline is summarized in Fig.~\ref{fig:autoencoder-example}. Anomaly detection refers to the practice of calculating a well defined `anomaly score' in order to determine if an event or group of events are significantly discrepant  from elements of the background distribution \cite{fraser2021challenges}. In statistical language, this process involves studying the asymptotic properties of the  \textit{anomaly function} $TS(x,y)$, which is generally a function of the image data inputs $x$ and CVAE outputs $y$. After data taking, one would calculate $TS$ for all data in the run. The resulting distribution can then be inspected for excesses or `anomalies' with respect to the simulated background-only distribution.

The function we adopt to perform the anomaly detection analyses is the following function that returns larger values when it is fed data that is more signal-like:
\begin{align}
\label{eqn:TS}
TS =  \left(\frac{1}{N}\sum_{i=0}^N(x_i-y_i)^2+\beta L_\text{KL}\,\right) + R\, H_B \;.
\end{align}
The term in parentheses is the CVAE loss function defined in Eq.~\ref{eqn:recon_loss} and the last term is a binary cross-entropy between a vector of zeros  and the output of the pre-trained CNN. The CNN was trained to classify ER background as 0 and NR as 1 (see Ref.~\cite{Khosa:2019qgp} and Fig.~\ref{fig:autoencoder-example} for architecture details). Hence, $H_B$ is larger in the presence of signal in the data set, e.g. when $p(x)\rightarrow 1$: 
\begin{align}
\label{eqn:cross_ent}
H_B = &-  \frac{1}{N}\sum_{i=0}^N\log \left(1-p\left(x_i\right)\right)\;,
\end{align}  
where $p(x)$ is understood as the output of the classifier given the input image $x_i$\footnote{We tuned the CNN architecture slightly with respect to Ref.~\cite{Khosa:2019qgp}, and after training on a larger dataset it achieved a classification accuracy of $\sim98\%$.}. The parameter $R$ in Eq.~\eqref{eqn:TS} scales the contribution of the cross-entropy term. In the limit $R\rightarrow 0$, we recover the reconstruction loss from Eq.~\eqref{eqn:recon_loss} for a purely unsupervised analysis, while for $R\rightarrow \infty$ we are effectively doing an supervised study\footnote{It should also be noted that while increased accuracy of this supervised classifier is useful,  the analysis  only focuses on TS defined in Eq.~\eqref{eqn:TS} and its distribution after taking  data.}. We note that it may seem natural to adopt the reconstruction loss as the anomaly function. However, we found that by generalizing it in this way, we utilize the power of a pre-trained CNN in a semi-supervised fashion which  dramatically improves the anomaly awareness, as will be discussed in Sec.~\ref{sec::op_nom}.

\subsection{Results for pure signal/background data}
\label{sec:signal}
\begin{figure}
\centering
\includegraphics[width=0.485\textwidth]{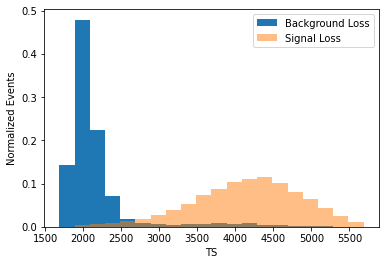}
\includegraphics[width=0.53\textwidth]{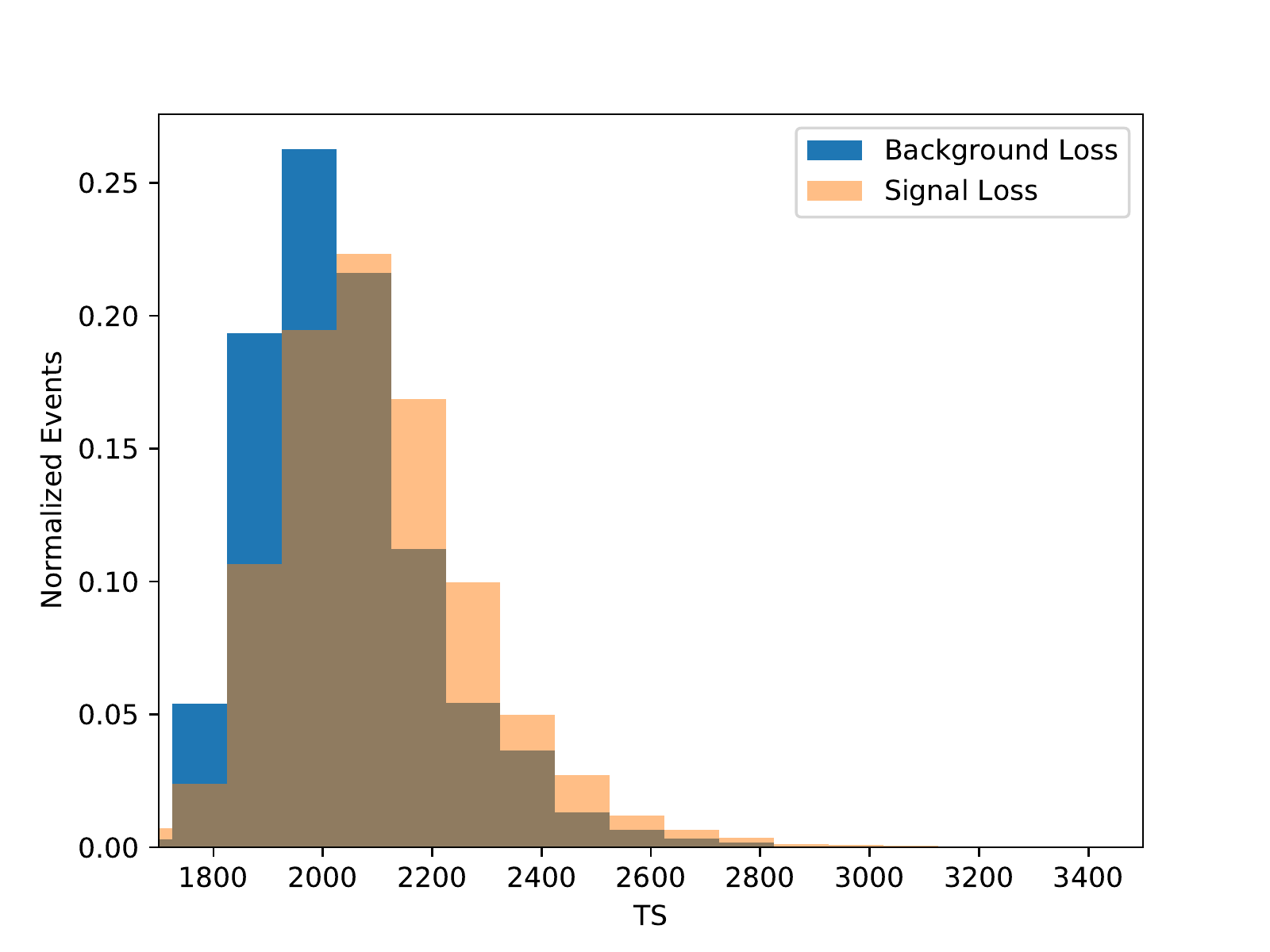}
\caption{\label{fig:cvae-signal-example} \textbf{Top:} Distribution of the anomaly function  defined in Eq.~\eqref{eqn:TS} for only ER (blue) events and only NR events (orange) and $R=170$. As expected, NR events are displayed as more `anomalous' with higher values of $TS$. \textbf{Bottom}: Same as top but for $R=0$. In this case there is a higher degree of overlap between the ER-only and NR-only distributions, and hence one would expect significantly weaker anomaly awareness. }
\end{figure}

We apply the now trained CVAE model to test data sets (data sets the networks were not trained on) consisting of $10000$ ER and NR (WIMP) events.  To assess the discrimination power of our technique, we histogram the anomaly function defined in Eq.~\eqref{eqn:TS} for pure signal injection (NR) vs background-only (ER) (all 10000 events). In other words, we bin TS without re-weighting the distributions based on cross-section and WIMP mass. We do this  for two cases: $R=0$ (bottom) and $R=170$ (top), the latter value being observed to provide optimal results, as will be discussed in Sec.~\ref{sec::op_nom}.  The results are normalized to unit area and are shown in Fig~\ref{fig:cvae-signal-example}. It can be seen that the   distribution of the WIMP events has more density at larger TS on average than the ER events for both values of $R$, implying that the model is learning something fundamental about the electron recoil events that is not present in the WIMP events.  It is directly obvious however, that the case of  $R=0$ (purely unsupervised) yields significantly worse anomaly awareness than the case of $R=170$ (semi-supervised).  

\subsection{Results for background plus signal data}
\label{sec:results}
\begin{figure}[t]
	\centering
	\includegraphics[scale=0.6]{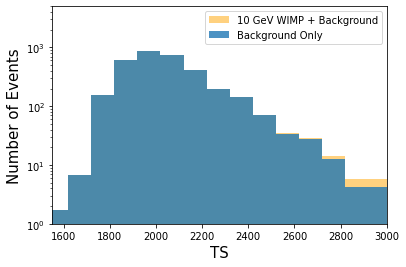}
	\includegraphics[scale=0.6]{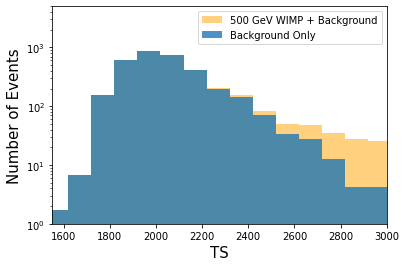}
	\includegraphics[scale=0.6]{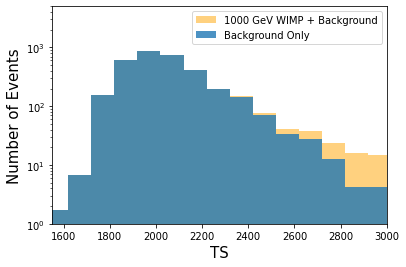}
	\caption{\label{fig:cvae-background-data} Simulated distributions of   TS after $5\,t\,y$ of exposure for the electron recoil (background) sample and the pseudo-data sample with an injected WIMP signal weighted to the relevant mass and cross-section for (from top to bottom) $10$ GeV, $500$ GeV and $1000$ GeV. The scattering cross-section used was $\sigma=10^{-45}$cm$^2$ and the $R$ value taken was $R=170$. The anomalous (signal) data is observed as an excess in the tail of the distribution.}
\end{figure}

The main analysis involves  applying the network to a sample of ``pseudo-data'', created by injecting the WIMP signal NR events into a dataset of ER events scaled to the appropriate normalization. The ER and WIMP events are weighted to an exposure of  $5\,t\,y$. Furthermore the WIMP signal is weighted to a spin-independent scattering cross-section of $\sigma_\text{SI} = 10^{-45}$ cm$^2$ for masses 10, 500 and 1000 GeV. In re-weighting the distributions in such a fashion we are effectively assuming we have used the Asimov data \cite{Cowan_2011}. One may note that the cross-section used is currently excluded for some DM mass values \cite{Benabderrahmane:2019yhk}. For the time being, we are simply demonstrating the novelty of the procedure which is exemplified for a relatively large cross-section. An analysis considering non-excluded regions of the DM parameter space is considered later in Sec.~\ref{sec:forecasting}.

 We expect to see a pseudo-data distribution that matches the electron recoil distribution to a high degree in the low TS  areas with excesses seen in tails of the distributions. This is presented in Fig.~\ref{fig:cvae-background-data} where we show the background-only as well as pseudo-data TS distributions for the three WIMP mass benchmarks. The excess is least visible for 10 GeV since this mass gives the lowest number of expected events (13) compared to 500 GeV (187) and 1000 GeV (95). The significance of the excess as a function of exposure as well as a comparison to the XENONnT forecasted $5\sigma$ discovery is presented in the next section.

 \subsection{Optimizing anomaly awareness}
\label{sec::op_nom}

\begin{figure}[t]
	\centering
	\includegraphics[scale=0.58]{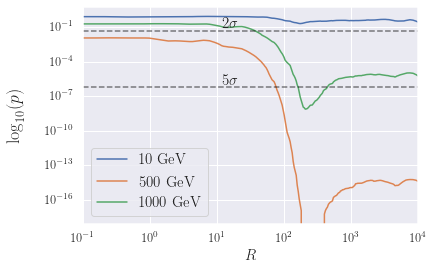}
	\caption{$p$-value to reject the background-only (ER) hypothesis as a function of the scale parameter $R$ using an exposure of $5\,t\,y$ and a spin-independent scattering cross-section of $\sigma=10^{-45}$cm$^2$. $R\rightarrow 0$ implies TS is calculated only using the CVAE (purely unsupervised), while $R\rightarrow\infty$ implies TS is dominated by the output of the CNN (purely supervised). The optimal value for $R$ is observed to be in the range $R\in[150-300]$, indicating the need for a semi-supervised approach for optimal anomaly awareness. Calculation of the $p$-value is detailed in Sec.~\ref{sec:forecasting}.  \label{fig:pvsR}}
\end{figure}

The probability with which we can reject the background-only hypothesis was calculated with a simple Pearson $\chi^2$ test:
\begin{align}
\chi^2_p = \sum_\text{bins} \, \frac{(TS_\text{ER}-TS_\text{ER+NR})^2}{TS_\text{ER}+TS_\text{ER+NR}}\;.
\end{align}
We assume Wilk's theorem holds such that $\chi^2_p$ follows a $\chi^2$-distribution with one degree of freedom. We expect this to hold given inspection of Fig.~\ref{fig:cvae-signal-example}, clearly displaying asymptotic forms of a nested hypothesis, with the null hypothesis (background-only) being realized explicitly when $R=0$. The $p$-value to reject the null (background-only) hypothesis is then given by  
\begin{align}
\label{eqn::pval}
p = 1-\text{CDF}({\chi^2_\text{1d.o.f}})\,,
\end{align}
where CDF is the cumulative density function of a $\chi^2$-distribution with one degree of freedom. 

 In  Fig.~\ref{fig:pvsR} we calculated the $p$-value for rejecting the background only hypothesis while scanning over values of $R$. We observe that for any given DM mass, a purely supervised anomaly function (large $R$) has better discriminating power than a purely unsupervised one ($R=0$), as expected, with $R\lesssim 10$ and $R \gtrsim 10^3$ yielding a constant $p$-value. Interestingly, however,  we find that there is a range for $R$ centred at about $R\sim 170$ for the cross section used, where the $p$-value is minimized. This appears to be the case regardless of the DM mass (as long as they are above tens of GeV).  This does not necessarily apply to varying the cross-section, however, as we will discuss in Sec.~\ref{sec:forecasting}.   Having the best background hypothesis rejection occur before the plateau at large $R$ indicates the need for both the CVAE and CNN to be utilized in the anomaly function.   This is due to the fact that in a purely supervised scenario, constructing the anomaly test in this way would yield significant loss of information due to (nearly) all ER events being binned at $TS=0$ and NR events being binned at some arbitrary, large  $TS$ value. The inclusion of the CVAE loss into the anomaly function serves to provide a distribution that captures all available information, effectively getting the best of both unsupervised and supervised approaches, qualitatively  increasing the statistical power of any observed excess in $TS$.  This demonstrates the overall improvement of our analysis by implementing a semi-supervised approach.
 
An additional thing to note is that, whilst we see that the anomaly function in Eq.~\eqref{eqn:TS} gives good results, one is technically free to choose any such function, as long as the same operation has been consistently done  to the background-only and pseudo-data sets. A comprehensive study of the different anomaly functions and their power in the context of semi-supervised learning in different beyond the standard model (BSM) searches is outside the scope of this work and is left for future study.

\subsection{Forecasting sensitivity}
\label{sec:forecasting}

When the data being used are a representation of how they would look after some time, the $p$-value in Eq.~\eqref{eqn::pval} can be referred to as the `projected sensitivity'. We note that a more formal statistical inference may be done, however, to demonstrate the novelty and generalizability of this method, we decided to keep things simple.
\begin{figure}
	\centering
	\includegraphics[scale=0.6]{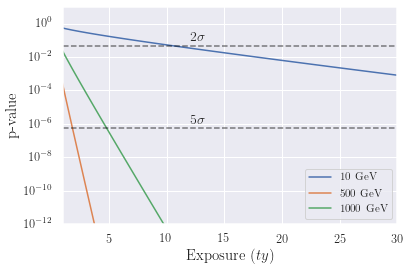}
	\includegraphics[scale=0.6]{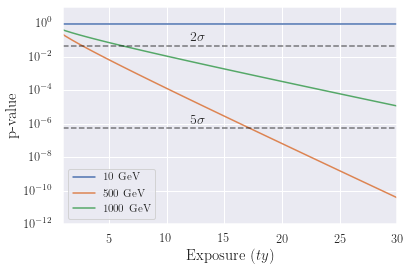}
	\caption{\textbf{Top:} Projected $p$-value of rejecting the background-only hypothesis vs exposure for a scattering cross-section of $\sigma=10^{-45}$ cm$^2$. \textbf{Bottom:} The same but for $\sigma=10^{-46}$ cm$^2$. The $R$ value taken was $R=170$. \label{fig:pval_vs_exp} }
\end{figure}
It is often useful to project or `forecast' what the $p$-value or `sensitivity' of discovery would be (assuming the WIMP exists) after some exposure. In Fig.~\ref{fig:pval_vs_exp}  we plot the $p$-value calculated as in Eq.~\ref{eqn::pval} as a function of exposure for three benchmark masses and a fixed value of $R=170$. For demonstration, we take a currently non-excluded cross section of $\sigma=10^{-45}$ cm$^2$ (top) as well as  $\sigma=10^{-46}$ cm$^2$ (bottom). We additionally include dashed lines representing $p$-values corresponding to a number of unit Gaussian standard deviations of $2\sigma$ and $5\sigma$. We can see that, as expected, the relatively low number of events for a 10 GeV makes it difficult to significantly detect an excess over the background-only distribution, however 500 GeV and 1000 GeV remain reasonably detectable.  
\begin{figure}
	\centering
	\includegraphics[scale=0.6]{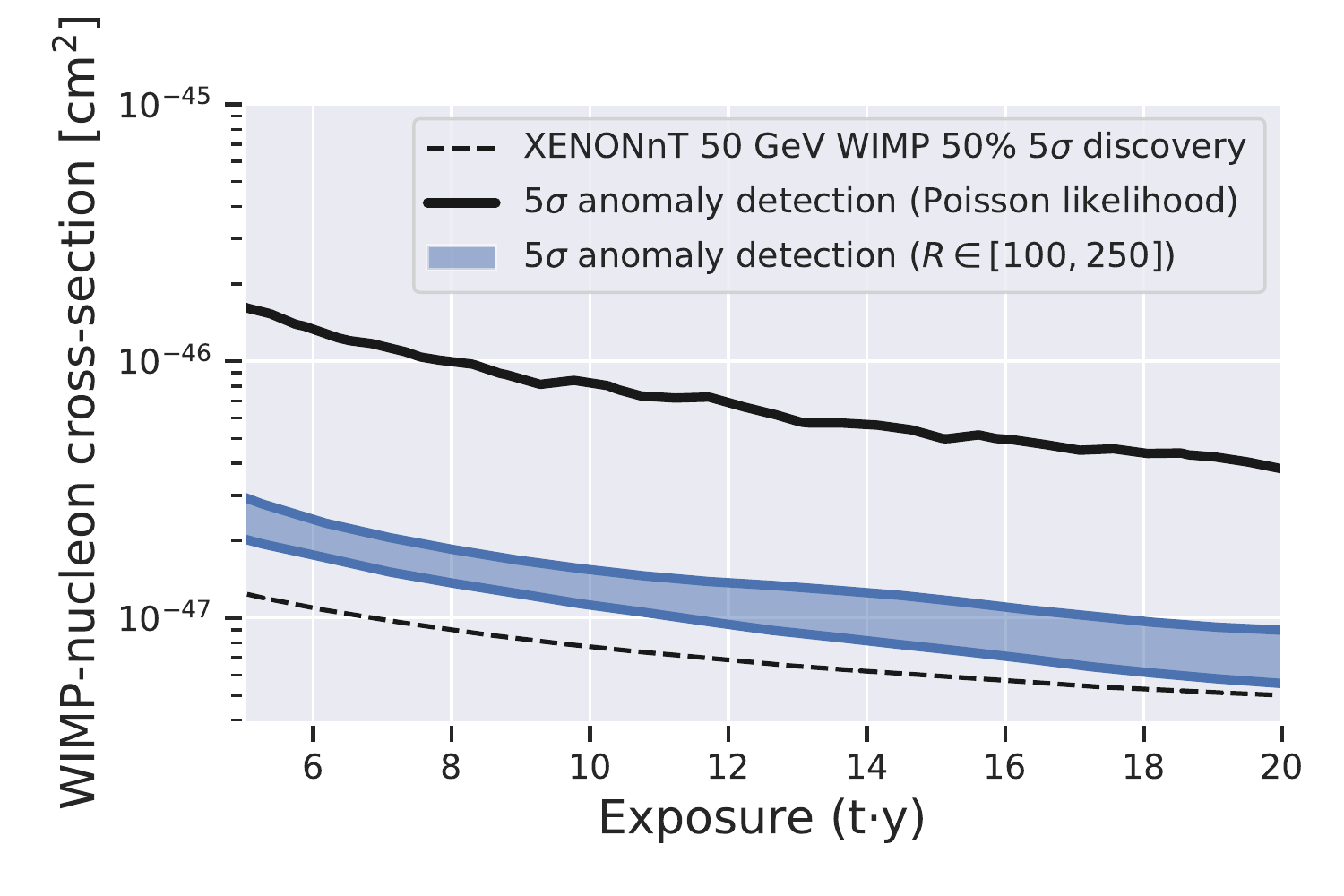}
\caption{Contours of $5\sigma$ forecasted discovery significance in the scattering cross-section vs. exposure plane assuming the presence of a 50 GeV WIMP. Blue bands  represent our semi-supervised anomaly detection conducted with a varying number of $R$ values within the range $R\in[100,250]$. The black solid line is the projected $5\sigma$ anomaly discovery sensitivity calculated using a simple traditional likelihood based approach as detailed in the text.   The black dashed line is the official XENONnT collaboration result for the smallest cross-sections at which the experiment would have a 50\% chance of observing a 50 GeV WIMP with significance greater than $5\sigma$.  \label{cs_vs_exp}}
\end{figure}

Finally,  we plot  the $5\sigma$ sensitivity contour in the cross-section/exposure plane for a variety of $R$ values in the range $R\in[100,250]$. We present the results as a band to illustrate how the sensitivity can maximally and minimally vary depending on the choice of $R$. In order to observe the strongest sensitivity, we do this for a $50$ GeV WIMP which is the mass which approximately gives the greatest expected number of events for any given cross-section. We show the result in  Fig.~\ref{cs_vs_exp}.  The variation of sensitivity given a choice of the $R$ parameter  exemplifies the need to use a range of $R$ values (or indeed, anomaly functions in general) in a real analysis, since the $R$ value did not enter the training process whatsoever and we are free to choose the one which optimizes anomaly detection (rejecting the background-only hypothesis).  In order to qualitatively gauge the performance of our semi-supervised anomaly detector relative to a traditional statistical approach, we estimate the $5\sigma$ anomaly sensitivity using a traditional frequentist approach by bootstrapping  samples of the corrected S1/S2 (cS1/cS2) distribution from \lb \, in order to calculate the asymptotic form of a -2ln(Poisson) distribution corresponding to the background-only and background+anomaly hypotheses (see Ref.~\cite{Cowan_2011} for details on the standard procedure). We present the forecasted $p-$value to reject the background-only hypothesis at $5\sigma$ in Fig.~\ref{cs_vs_exp} as a black contour. We observe that  our semi-supervised anomaly detector indeed  performs better than the likelihood based method by about almost an order of magnitude. We expect this is the case since for classical  statistical tests, the choice of likelihood function is paramount in determining the strength of the statistical test. Hence, if one could find the `correct' likelihood function that exactly describes how nature interacts with the experiment, this would be the optimal choice. Of course, usually we do not have access to such a function in practice, and so we usually take an educated guess (Poisson likelihoods in particle physics for example). What our machine learning method has essentially done is, based on simulations that we consider accurate representations of the physics, `learn' the underlying properties of the data that would give the `best estimate' of the likelihood function, even if such an analytical function doesn't exist. This is the general premise of simulation based inference (SBI) (see Ref.~\cite{brehmer2020simulationbased} for a review). A more detailed study of the general improvement to modern SBI and anomaly detection methods by utilizing multiple, separately trained  supervised/unsupervised architectures is beyond the scope and is left for future work.  

 For completeness, as a dashed black line in Fig.~\ref{cs_vs_exp}  we also include the official XENONnT projected $5\sigma$ discovery sensitivity from  Ref.~\cite{XENON:2020kmp}, which denotes  the smallest cross-sections at which the experiment would have a 50\% chance of observing an excess with significance greater than $5\sigma$ corresponding to a WIMP with mass  50 GeV. It should be noted that this contour should not be directly compared to our analyses, since it explicitly refers to the discovery potential of a 50 GeV WIMP, which comprehensively involves the profiling and minimization over model parameters in the likelihood analysis, as well as the inclusion of several complicated detector dependent terms. The result is that such a sensitivity will be much stronger than that of any anomaly search.  It is also important to note that the CVAE was trained on background events \textit{only}. That is, \textit{any} events that do not look like background will show up as having an anomalous TS distribution. Whilst the projected sensitivities we present in this section implicitly depend on the WIMP model (mass/cross-section) as we needed to re-weight the $TS$ distributions, the general principle is completely model independent in that one may always run a real data set through the CVAE-anomaly detector to search for anomalous physics of any kind. We thus encourage these methods to be adapted to other low background searches for new physics, especially with detectors for which more than one DM model may be probed.

Regardless, it is qualitatively useful to see how the semi-supervised learning approach compares to statistical approaches, even though it is given no direct information about the WIMP model parameters.  With further optimization, as well as access to other auxiliary data, it may indeed be very possible to train such an anomaly detector to surpass the currently used state of the art statistical forecasting methods.

\section{Conclusion}
\label{sec:conc}

In this study we present an example of a model-independent search for anomalous DM direct detection events using a convolutional variational autoencoder (CVAE). The XENONnT experiment was used as a test-bed for this methodology wherein a set of composite images including the detector response and S1/S2 channels were used as input data. The CVAE was trained on a set of electron recoil background events, learning to encode the input images onto a lower dimensional latent space, then decoding from this latent space back to the original image. The reconstruction loss is defined as the difference between the pixel information of the original image and the decoded image. The loss is thus a proxy for anomalousness as images that do not conform to the general behaviour and distribution of background events will inherently produce higher reconstruction losses as the network has not learned the properties of these events. 

Additionally, in this work we exploit a pre-trained supervised convolution neural network which boosts the anomaly detection power of the CVAE. Interestingly, this means that the best performance for anomaly detection is found for a combination of both unsupervised and supervised networks, in a semi-supervised fashion. We parameterize the influence of the supervised part with a hyperparameter $R$. We find that the optimal $R$-value to use is invariant with respect to the DM mass, however, we emphasize that in a real analysis, detailed consideration  as well as optimization of any hyperparameters (e.g. $R$ values) should be undertaken.
 
With respect to future sensitivity, we find that the projected $5\sigma$ anomaly sensitivity corresponding to a WIMP mass of 50 GeV outperforms that of a simple classical likelihood based approach.  We also find that our anomaly detector performs similarly, although as expected, slightly weaker (e.g. $\sim$ a factor of a few), than the official prediction of the XENONnT Collaboration for the discovery of a 50 GeV WIMP. Even though these results are not exactly statistically comparable,  this is an encouraging result which may render interesting for the Collaboration to run this type of analysis given their access to additional data and more sophisticated experimental techniques.

Finally, we would like to point out that, although a given particle physics interaction was used (point-like, spin-independent, elastic) to train the supervised network, the analysis is quite agnostic regarding the underlying DM model which is generating the excess of anomalous events, whose main characteristic is consisting of nuclear recoils. Given the absence of any positive signals in the last decades, trying to be as agnostic as possible regarding which signals we expect to detect seems the right approach.

\section*{Acknowledgements}
 The authors would like to thank 
 Ver\'onica Sanz, Bryan Zald\'ivar and Gabriel Maicas for useful discussions.
JHG is supported by the \emph{Generalitat Valenciana} through the GenT Excellence Program (CIDEGENT/2020/020), and the Ministerio de Ciencia e Innovacion MICIN/AEI (10.13039/501100011033) grants PID2020-113334GB-I00 and PID2020-113644GB-I00. AS acknowledges support from the research grant “The Dark Universe: A Synergic Multimessenger Approach” No. 2017X7X85K funded by MIUR and the project “Theoretical Astroparticle Physics (TAsP)” funded by the INFN. Part of this work was done using the Artemisa computing infraestructure at IFIC, co-funded by the European Union through the 2014-2020 FEDER Operative Programme of Comunitat Valenciana, project IDIFEDER/2018/048.

\bibliographystyle{my-h-physrev}
\bibliography{DLDD_unsup.bib}

\end{document}